%% file: mhd_gap.tex
\documentclass[fleqn,usenatbib]{mnras}
\input{preamble.tex}

\graphicspath{{./}{figures/}}

\title[Gap Opening in MHD Protoplanetary Disks]{Gap Opening in Protoplanetary Disks: 
Gas Dynamics from Global Axisymmetric Non-ideal MHD Simulations with Consistent Thermochemistry}

\author[X. Hu et al.]{
Xiao Hu (\cntextsc{胡晓})$^{1,2}$\thanks{E-mail: xiao.hu.astro@gmail.com}\orcidlink{0000-0003-3201-4549},
Zhi-Yun Li$^{1}$,
Lile Wang (\cntextsc{王力乐})$^{3}$\thanks{E-mail: lilew@pku.edu.cn}\orcidlink{0000-0002-6540-7042},
\newauthor
Zhaohuan Zhu (\tctext{朱照寰})$^{4,5}$ \orcidlink{0000-0003-3616-6822}
and Jaehan Bae$^{2}$\orcidlink{0000-0001-7258-770X}
\\
$^{1}$Department of Astronomy, University of Virginia, Charlottesville, VA 22904, USA\\
$^{2}$Department of Astronomy, University of Florida, Gainesville, FL 32608, USA\\
$^{3}$The Kavli Institute for Astronomy and Astrophysics, 
Peking University, Beijing 100084, China\\
$^{4}$Department of Physics and Astronomy, University of
  Nevada, Las Vegas, 4505 S. Maryland Parkway, Las Vegas,
  NV 89154, USA\\
$^{5}$Nevada Center for Astrophysics, University of Nevada, Las Vegas, 4505 South Maryland Parkway, Las Vegas, NV 89154, USA\\
}

\date{Accepted XXX. Received YYY; in original form ZZZ}

\pubyear{2023}

\begin{document}
\label{firstpage}
\pagerange{\pageref{firstpage}--\pageref{lastpage}}
\maketitle

\begin{abstract}
Recent high angular resolution ALMA observations have revealed numerous gaps in protoplanetary disks. A popular interpretation has been that planets open them. Most previous investigations of planet gap-opening have concentrated on viscous disks. Here, we carry out 2D (axisymmetric) global simulations of gap opening by a planet in a wind-launching non-ideal MHD disk with consistent thermochemistry. We find a strong concentration of poloidal magnetic flux in the planet-opened gap, where the gas dynamics are magnetically dominated. The magnetic field also drives a fast (nearly sonic) meridional gas circulation in the denser disk regions near the inner and outer edges of the gap, which may be observable through high-resolution molecular line observations. The gap is more ionized than its denser surrounding regions, with a better magnetic field-matter coupling. In particular, it has a much higher abundance of molecular ion HCO$^+$, consistent with ALMA observations of the well-studied AS 209 protoplanetary disk that has prominent gaps and fast meridional motions reaching the local sound speed. Finally, we provide fitting formulae for the ambipolar and Ohmic diffusivities as a function of the disk local density, which can be used for future 3D simulations of planet gap-opening in non-ideal MHD disks where thermochemistry is too computationally expensive to evolve self-consistently with the magneto-hydrodynamics. 
\end{abstract}



\section{Introduction} 
\label{sec:intro}
Planet gap opening in protoplanetary disks (PPDs hereafter) is a classic research topic in planet formation \citep{1993prpl.conf..749L,2010apf..book.....A}. The gravity from the planet launches spiral density waves, which deposit significant torque in the disk \citep{1979ApJ...233..857G,2012ARA&A..50..211K}. This torque is large enough to alter the disk structure, resulting in a local low-density annulus (i.e., a planetary gap) along the planetary orbit \citep{2006Icar..181..587C,2013ApJ...769...41D}. This process affects both the accretion of gas and solids and the orbital migration of the planet \citep{2014prpl.conf..667B,2015ApJ...806L..15K}. These factors largely shape the outcome of an exoplanet system, determining planet mass and orbital radius \citep{2004ApJ...616..567I,2012A&A...547A.112M}.

The critical mass of gap opening can be derived from the shear stability of steep pressure gradient \citep{1993prpl.conf..749L} and the time scale for the disk viscosity to refill it \citep[e.g.,][]{2013ApJ...769...41D,2018A&A...612A..30B,2018ApJ...857...20H}. The process of heat dissipation is important for the
disk torque  \citep{2014MNRAS.440..683L,2017MNRAS.472.4204M}.
The gap depth scaling relations were obtained in semi-analytical work and simplified 2D (infinitely thin) disk simulations, and they were shown to hold well when moving to higher dimensions  \citep{2016ApJ...832..105F}. Three-dimensional (3D)  simulations provide, in addition, a characterization of the meridional flows, which turn out to be amenable to ALMA observations and thus provide a powerful tool to probe the planet-disk interaction \citep[e.g.,][]{2019Natur.574..378T,2020ApJ...890L...9P}.

It is well known that disk dynamics and evolution are strongly affected by magnetic fields. For example, magnetic fields have been shown to play important roles in carving gaps by direct surface accretion stream \citep{2017MNRAS.468.3850S} or radial magnetic flux redistribution \citep[e.g.,][]{2018MNRAS.477.1239S,2019ApJ...885...36H, 2021ApJ...913..133H,2021MNRAS.507.1106C} in protoplanetary disks, and maintain meridional flows \citep{2022MNRAS.516.2006H}. In addition, they likely dominate the angular momentum transport in PPDs through magnetized disk winds \citep{1982MNRAS.199..883B,2013ApJ...769...76B}. Their dynamical importance motivated numerical studies that combine planet-disk interaction with magnetic fields
and wind-driven accretion. Early simulations along this line usually assumed ideal MHD and incorporated only a toroidal magnetic field without launching a disk wind \citep{2003MNRAS.339..993N,2003ApJ...589..543W,2011A&A...533A..84B}. More recent local shearing-box MHD simulations have included a net poloidal magnetic flux and found that the magnetic flux
gets concentrated into the planet-induced gap, making
the gap deeper and wider due to enhanced MRI turbulence within the gap \citep{2013ApJ...768..143Z,2017MNRAS.472.3277C}. However, besides being local, these studies are typically unstratified in the vertical direction and thus incapable of wind-driven angular momentum transport.
As importantly, disks are weakly magnetized, so non-ideal MHD effects are important. \citet{2013ApJ...779...59G} included Ohmic dissipation in a global simulation, but its extent is limited to 4.5 disk scale heights, limiting the disk wind's treatment. In this paper, we will go beyond the previous work by extending the simulation domain to close to the polar axis and including ambipolar diffusion to have comprehensive coverage of the wind region,  which is important for the scale of tens to hundreds of au most accessible to ALMA observations. 

Because of the complexity and especially the computational demand of self-consistent thermochemistry evolution, we will carry out our global non-ideal MHD simulations assuming axisymmetry as a first step, adopting a torque profile to approximate the planet-disk interaction. This approximation will be relaxed in future 3D simulations. The focus is on the gas kinematics in the vicinity of the planet-opened gap, which is starting to be probed in increasing detail by ALMA. Our approach complements the recent work of \cite{2023ApJ...946....5A}, who carried out 3D simulations of gap-opening by planets in non-ideal MHD disks but with a prescribed spatially-uniform ambipolar Elsasser number inside the disk. However, a significant spatial variation of ambipolar Elsasser number is expected given the large dynamic range of the density involved in the gap opening process. We seek to capture this variation through consistent thermochemistry that includes ionization.

The paper is organized as follows. In \S\ref{sec:methods}, we describe the numerical methods and simulation setup, especially the planet torque implementation and validation through hydro-only setups. \S\ref{sec:model-fiducial} analyzes the results from our simulation, focusing on the magnetic effects on the gas kinematics in and around the planet-opened gap and the relation between the ambipolar and ohmic diffusivities and the disk density. 
We discuss the observation implications of our results and conclude in \S\ref{sec:discussion}.

\section{Methods}
\label{sec:methods}
\subsection{Disk}
We simulate the disk evolution with the combination of non-ideal MHD effects using the higher-order Godunov MHD code \verb|Athena++| \citep{2020ApJS..249....4S}, ray-tracing radiative transfer for high-energy photons, and consistent thermochemistry. For each MHD timestep, the non-equilibrium thermochemistry is co-evolved in each zone throughout the simulation domain with a semi-implicit method. In general, this numerical system is almost the same as \citet{2019ApJ...874...90W} (WBG19 hereafter). We refer the reader to WBG19 and references therein for the details of the initial conditions and the overall setup of thermochemical reactions, 
which included 28 species: \neg{e}
(free electrons), \pos{H}, H, \chem{H_2}, He, \pos{He}, O, \pos{O}, OH, \pos{OH},
\chem{H_2O}, C, \pos{C}, CO, CH, \pos{CH}, \pos{HCO}, Si, \pos{Si}, SiO, \pos{SiO}, \pos{SiOH}, S, \pos{S}, \pos{HS}, Gr, \pos{Gr}, \neg{Gr}. Here Gr and Gr$^\pm$ denote neutral
and singly-charged dust grains, respectively.
For the boundary conditions, we inherited the setups in \citet{2021ApJ...913..133H}, i.e., similar to WBG19 except
for the toroidal field above the disk region (viz. inside
the wind region) at the inner radial boundary: we set
$B_{\phi} = -B_r|_{t=0}$ (the initial value of $r$ component)
to suppress magnetic instabilities there. 
Other hydrodynamic and field components are identical to WBG19. 

The dust grains are treated as single-sized
carbonaceous grains co-moving with the gas. The size is 
$a_\Gr = 5$ \ang. The basic properties of our
model are summarized in
Table~\ref{table:fiducial-model}. The main parameter that determines the dust thermochemical properties, dust grain cross section per hydrogen nuclei
\sigperH, is set to $8\times10^{-23}~\cm^2$
which corresponds to a dust-to-gas mass ratio of $7\times10^{-6}$. This is also the maximum \sigperH~in \citet{2021ApJ...913..133H}, which is 10 times lower than WBG19. 
\begin{table}
  \caption{Properties of Disk Model (\S\ref{sec:model-fiducial}) 
    \label{table:fiducial-model}
  }
  \centering
  \scriptsize
\begin{tabular}{l r}
\hline
  Item & Value\\
  \hline
  Radial domain & $2~\au \le r \le \ 100~\au$\\
  Latitudinal domain & $0.06~{\rm rad}\le\theta\le \pi/2~{\rm rad}$ \\
  Resolution & $N_{\log r} = 480$, $N_\theta= 128$ \\
  \\
  Stellar mass & $1.0~M_\odot$ \\
  \\[2pt]
  Initial mid-plane density &
  $8\times 10^{14}(R/\au)^{-2.2218}~m_p~\cm^{-3}$ \\[2pt]
  Initial mid-plane plasma $\beta$ & $10^4$ \\
  Initial mid-plane temperature &
  $305(R/\au)^{-0.57}~\K$ \\
  Artificial heating profile$\dagger$ &
  $305(R/\au)^{-0.57}~\K$ \\
  \\
  Luminosities [photon~$\s^{-1}$] & \\[5pt]
  $7~\eV$ (``soft'' FUV)  & $4.5\times 10^{42}$ \\
  $12~\eV$ (LW) & $1.6\times 10^{40}$ \\
  $3~\keV$ (X-ray) & $1.1\times 10^{38}$ \\
  \\
  Initial abundances [$n_{\chem{X}}/n_{\chem{H}}$] & \\[5pt]
  \chem{H_2} & 0.5\\
  He & 0.1\\
  \chem{H_2O} & $1.8 \times 10^{-4}$\\
  CO & $1.4 \times 10^{-4}$\\
  S  & $2.8 \times 10^{-5}$\\
  SiO & $1.7 \times 10^{-6}$\\
  \\
  Dust/PAH properties & \\
  $a_\Gr$ & $5$ \AA \\
  $\sigma_\Gr/\chem{H}$ & $8\times 10^{-23}~{\rm cm^2}$ \\
  \hline
  \end{tabular}
  \smallskip
\begin{tabular}{p{0.9\linewidth}}
$\dagger$: The artificial heating profile indicates the temperature of dusts in the mid-plane. Similar to WBG19, because the radiative transfer of diffuse infrared radiation field is not calculated in this paper, we adopt this profile of dust temperature as the floor of dust temperature.\\
\end{tabular}
\end{table}

\subsection{Magnetic Diffusion}

In a weakly ionized protoplanetary disk, the equation of motion for charged species ($\e^-$, ions and charged grains $\chem{Gr}^\pm$) is set by the balance between the Lorentz force and the neutral-ion drag force: 
\begin{equation}
  Z_je\left(\bm{E}^{'}
    + \frac{\bm{v}_j}{c}\times \bm{B}\right)
   = \gamma_j \rho m_j \bm{v}_j
\end{equation}
where, for a given charged particle type (denoted by $j$), $Z_j$ represents its charge number (in units of charge $e$), $m_j$ is its mass, and $\bm{v}_j$ is its drift velocity relative to the neutral background. $\gamma_j$ is defined as $\mean{\sigma v}/(m+m_j)$, where $m$ is the average particle mass of the neutrals, and $\mean{\sigma v_j}$ is the rate coefficient for momentum transfer between the charged particle and neutrals. The electric field in the frame moving with the neutrals is represented by $\bm{E}^{'}$.

For very small grains, the interaction between charged grains and neutral molecules is influenced by the electric field from induced electric dipoles in the neutrals. This $r^{-4}$ electric potential results in a temperature-independent collision rate coefficient $\langle\sigma v\rangle_j$\citep[e.g.,][]{2011piim.book.....D}. As grains grow larger, their geometrical cross-section becomes more significant in interactions with neutrals. This grain size-driven transition is reflected in the Hall parameter, which is the ratio of charged particles' gyrofrequency under the Lorentz force to their collision frequency with neutrals:
\begin{equation}
    \beta_j=\frac{\abs{Z_j}e B}{m_j c}\frac{1}{\gamma_j\rho}
    \label{eq:beta_j}
\end{equation}
In our simulation, we adopt the recipes in \citet
{2011ApJ...739...50B, 2014ApJ...791...72B} to calculate the collision coefficients between the charged grains and neutrals:
\begin{equation}
\begin{split}
  \langle\sigma v\rangle_\Gr = \rm{max}&\bigg[1.3\times10^{-9}\abs{Z_\Gr},\\
  4\times10^{-3}&\left(\frac{a_\Gr}{1\rm{\mu m}}\right)^{2}\left(\frac{T}{100\rm{K}}\right)^{1/2}\bigg]\rm{~cm^3~s^{-1}}
\end{split}
    \label{eq:sig_v}
\end{equation}
So the transition from the electric potential-dominated cross-section to the geometric cross-section is at $\sim nm$ scale. Given $T$=100K, any single charged grain with $a_\Gr > 5.7\times10^{-8}~\cm$ needs to consider the geometric effect when calculating the collision coefficient. This affects the Ohmic, Hall and Pederson conductivities:
\begin{equation}
\begin{split}
  & \sigma_\O=\frac{ec}{B}\sum_j n_j|Z_j|\beta_j\ ,
  \\
  & \sigma_\H=\frac{ec}{B}\sum_j\frac{n_j Z_j}{1+\beta_j^2}
  \ ,\\
  & \sigma_\P=\frac{ec}{B}\sum_j\frac{n_j
    |Z_j|\beta_j}{1+\beta_j^2}\ .
\end{split}
\label{eq:sigfull}
\end{equation}
Here the summation index $j$ runs through all charged
species, with $n_j$ and $Z_j$ donating, respectively, the number density and charge of individual charged species. Using these conductivities, the general expressions for Ohmic diffusivity
$\eta_\O$, Hall diffusivity $\eta_\H$ and Ambipolar diffusivity $\eta_\A$ -- are \citep{2011ApJ...739...50B,2019ApJ...874...90W}.
\begin{equation}
\begin{split}
  &\eta_\O=\frac{c^2}{4\pi}\frac{1}{\sigma_\O}\ ,\quad
  \eta_\H=\frac{c^2}{4\pi}
  \frac{\sigma_H}{\sigma_\H^2+\sigma_\P^2}\ ,\\
  &
  \eta_\A=\frac{c^2}{4\pi}
  \frac{\sigma_\P}{\sigma_\H^2+\sigma_\P^2}-\eta_\O\ ,
\end{split}
\label{eq:diffu}
\end{equation}

The non-ideal induction equation is:
\begin{eqnarray}
\frac{\partial {\bm B}}{\partial t}=\nabla \times \left({\bm v}\times {\bm B}\right)
-\frac{4\pi}{c}\nabla \times ( \eta_\mathrm{O} {\bm J} \\+ \eta_\mathrm{H} {\bm J}\times {\bm b} + 
\eta_\mathrm{A} {\bm J}_{\bot}),\nonumber
\label{eq:induction}
\end{eqnarray}
where $\bm v$ is gas velocity, $\bm B$ is magnetic field, ${\bm b} =
{\bm B}/|B|$ is the unit vector representing field line direction.  
$\bm J$ is current density vector, and $\bm J_{\bot}$ is the current component perpendicular to the magnetic field. Note that the ambipolar diffusion and Ohmic dissipation terms are included in our non-ideal MHD simulations but not the Hall term. 

\subsection{Planet Torque}

\begin{figure}
    \centering
    \includegraphics[width=0.45\textwidth]{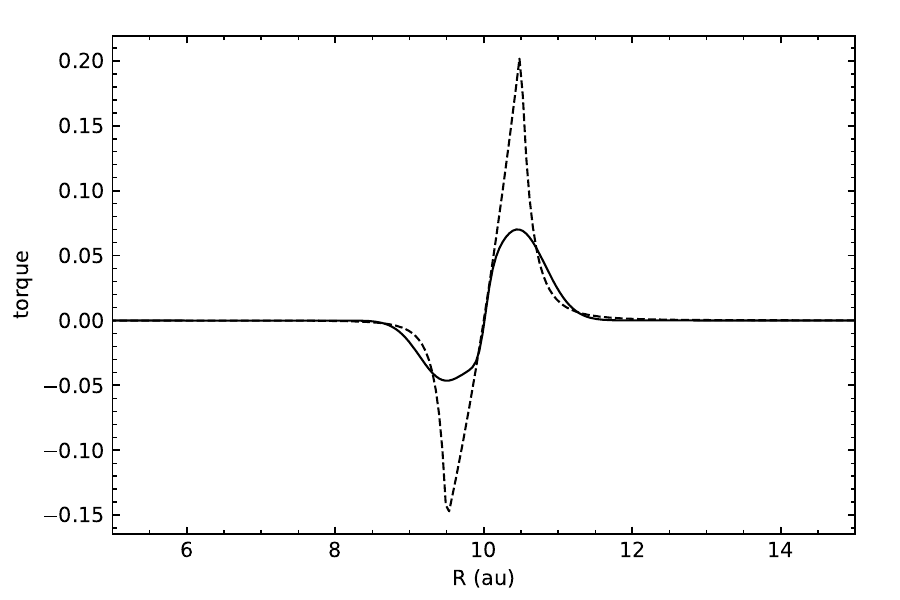}
    \caption{Torque profile (solid) described by Eq.~\ref{eq:Func_F}, with the impulse approximation \citep{1986ApJ...309..846L} plotted as dashed line for comparison.}
    \label{fig:tqprofile}
\end{figure}

We use the same approach as \citet{2017MNRAS.469.3813H} to replace the gap-forming planet with a gap-forming one-dimensional
torque density distribution across the two-dimensional disk along the cylindrical radial direction. The torque density profile is from \citet{2010ApJ...724..730D}:
\begin{equation}
    \Lambda = -\mathcal{F}\left(x,\beta,\zeta\right)\Omega^2R_{0}^2q^2\left(\frac{R_0}{H}\right)^4,
\end{equation}
where $R_0$ is the planet's orbital radius, $q=0.001$ is the planet-star mass ratio, $\Omega$ is the orbital angular velocity, $H$ is the disk scale height at $R_0$, $x = (R-R_0)/H$, $\mathcal{F}$ is a dimensionless function and $\beta$ and $\zeta$ are the surface density and temperature radial gradients (with a factor of -1) respectively. $\mathcal{F}$ is found by fitting the results of three-dimensional simulations \citep{2010ApJ...724..730D}:

\begin{multline}\label{eq:Func_F}
	\mathcal{F}(x,\beta,\zeta)= \left(p_{1}e^{\left(-\frac{(x+p_{2})^2}{p_{3}^2}\right)}+ p_{4}e^{\left(-\frac{(x-p_{5})^2}{p_{6}^2}\right)}\right) \\ \times 
	\textrm{tanh}(p_{7}-p_{8}x),
\end{multline}
where the fitting parameters $p_{1}$ through $p_{8}$ are constant for a given $\beta$, $\zeta$. We chose $\beta=0.5$ and $\zeta=1$ from the table 1 of \citet{2010ApJ...724..730D}, which are the closest fit to our disk parameter ($\beta=0.57$ and $\zeta=1$). We list the fitting parameters in Table~\ref{tab:p_values}. The torque profile in the arbitrary units is shown in Figure~\ref{fig:tqprofile}, together with the impulse approximation from \citet{1986ApJ...309..846L}. Note the torque profile here is for the disk midplane; for anywhere above the midplane, the strength of the torque is multiplied by a height factor $(R-R_0)^2/((R-R_0)^2+z^2)$ to approximate the effect of decreasing gravity away from the planet, with $z$ being the height above the midplane.

\begin{table}
\caption{Values of the parameter p in Eq.~\ref{eq:Func_F}.} 
\centering
\begin{tabular}{c c}
\hline
$p_n$ & Value \\
\hline
$p_1$ & $0.0297597$ \\
$p_2$ & $1.09770$ \\
$p_3$ & $0.938567$ \\
$p_4$ & $0.0421186$ \\
$p_5$ & $0.902328$ \\
$p_6$ & $1.03579$ \\
$p_7$ & $0.0981183$ \\
$p_8$ & $4.68108$ \\
\hline
\end{tabular}
\label{tab:p_values}

\end{table}

\begin{figure}
    \centering
    \includegraphics[width=0.45\textwidth]{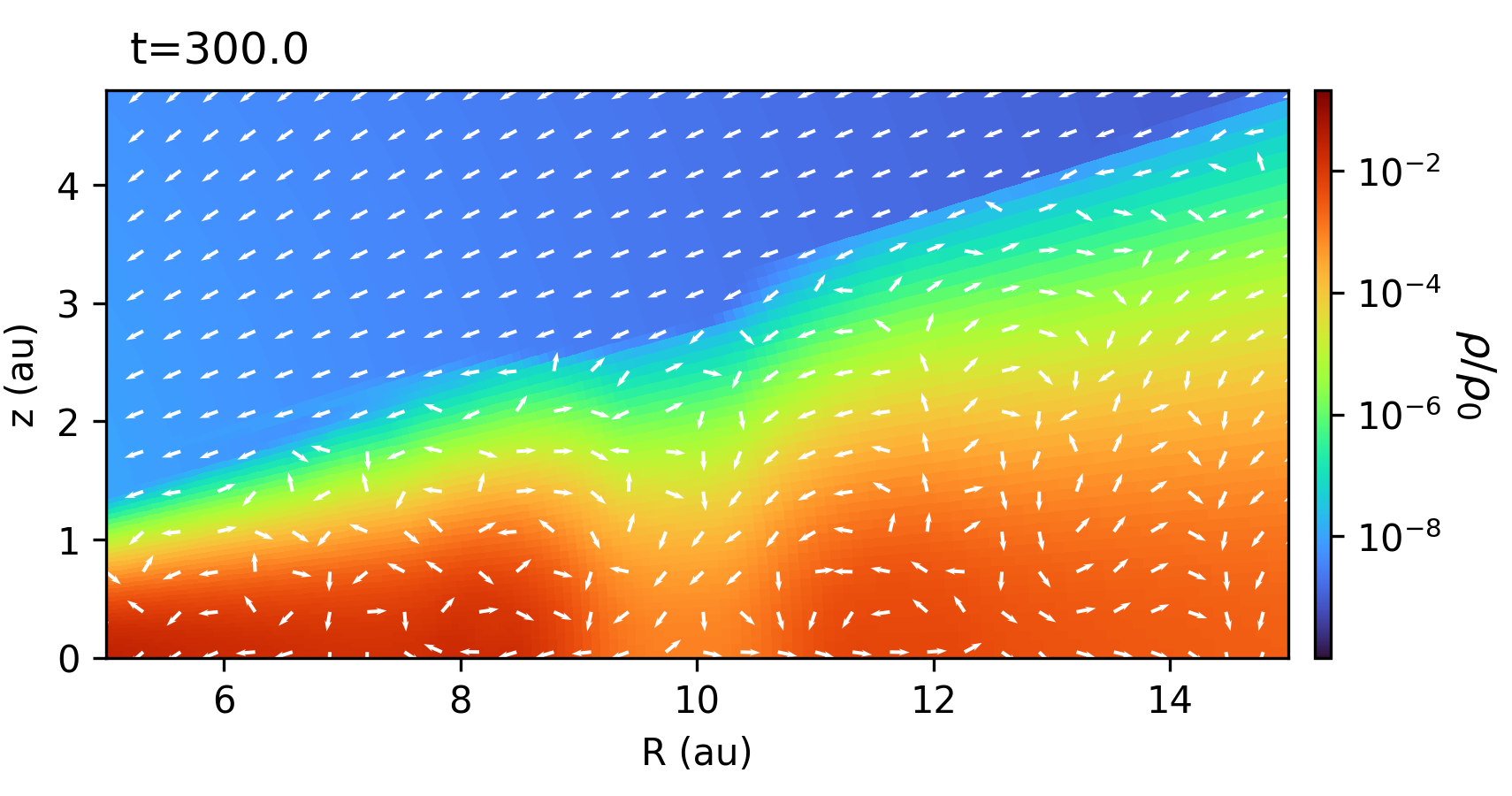}
    \caption{Test of planetary torque in a hydro-only simulation. Plotted are the density distribution (color map) and velocity field (vectors) near the planet-opened gap, showing a meridional circulation pattern with flows moving away from the planet (located at 10~au) near the midplane and collapsing flows towards the planet at higher altitudes.}
    \label{fig:hydro_vector}
\end{figure}

This profile is initially tested in a 2.5D axisymmetric inviscid hydro-only simulation that employs the same grid structure as our non-ideal MHD model, which we will describe later. In Figure~\ref{fig:hydro_vector}, we show that the adopted torque profile can produce a similar gas velocity pattern
as 3D hydrodynamics gap opening simulations in the meridional plane. In particular, the planet drives a dense midplane flow away from it near the midplane. At higher altitudes, steep density gradients drive a flow into the gap towards the planet, forming the characteristic ``collapsing flow" evident in 3D hydro simulations \citep[e.g.,][]{2016ApJ...832..105F}.  

\section{Model Results}
\label{sec:model-fiducial}

\begin{figure}
    \centering
    \includegraphics[width=0.45\textwidth]{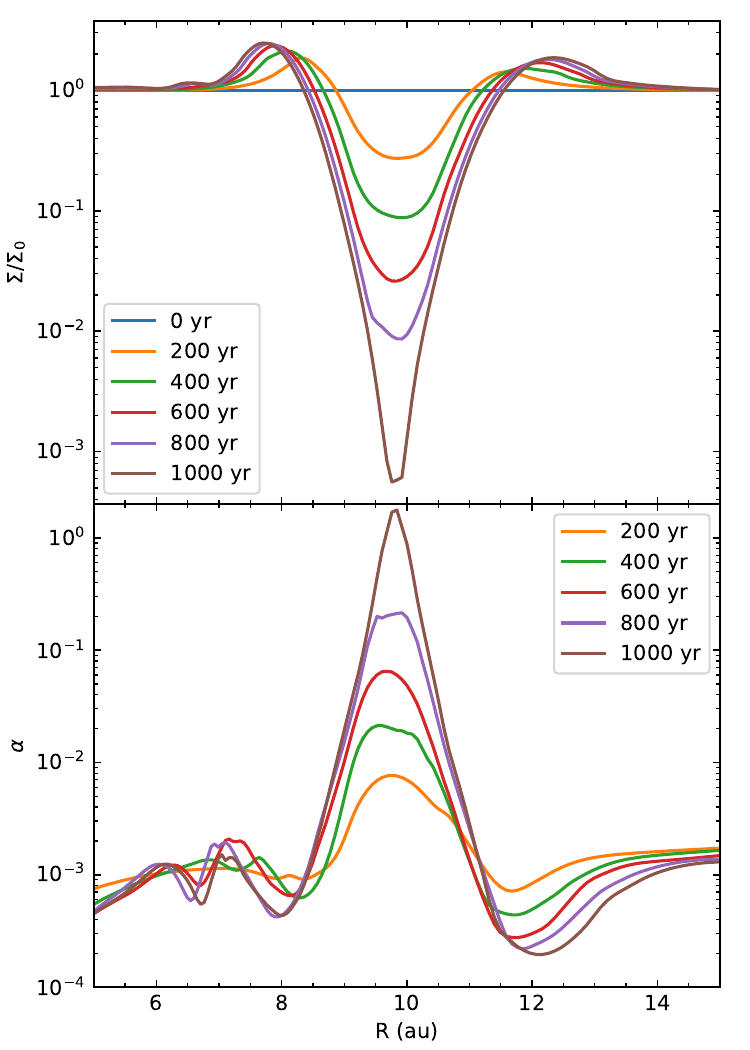}
    \caption{Gap opening as illustrated by the evolution of the surface density distribution over the first 1000 years (panel a). Panel (b) plots the ``effective" $\alpha$ parameter corresponding to the magnetic stresses, which are particularly important inside the low-density gap. 
    }
    \label{fig:sig1devo}
\end{figure}

We run our non-ideal MHD simulation to a time of $t=1000$~years, corresponding to about 30 planet's orbital periods at $10$~au. As expected, a gap is gradually opened by the planetary torque\footnote{A caveat of our planet's torque profile is a potential pile-up of material near the gap edges since it lacks spiral shocks from a real gap-opening planet that may transport angular momentum farther from the planet. Since this study mainly focuses on gas kinematics instead of density profile, we consider the order of unity difference of gas surface density near the gap edge acceptable.}, as illustrated in Fig.~\ref{fig:sig1devo}, where the column densities are plotted at different times (panel a), together with the effective $\alpha$ parameter corresponding to the magnetic stresses (panel b). The effective Shakura-Sunyaev
$\alpha$ is defined as the $R-\phi$ stress to pressure ratio:
\begin{equation}
\alpha_{int}=\left|\left.\int_0^{z+}{B_RB_\phi}dz\right|  \middle/\int_0^{z+}P_{\rm gas}dz\right.
\end{equation}
where the integral goes from the disk midplane to the upper disk surface ($z^+ \sim~ 4h$). The large values of $\alpha$ inside the gap point to the dynamical importance of the magnetic field, as discussed in detail later. 
Most of our analysis will be based on the snapshot at $t=800$~years; some analysis will be conducted at an earlier time $t=483$~years. 

\begin{figure*}
    \centering
    \includegraphics[width=1.0\textwidth]{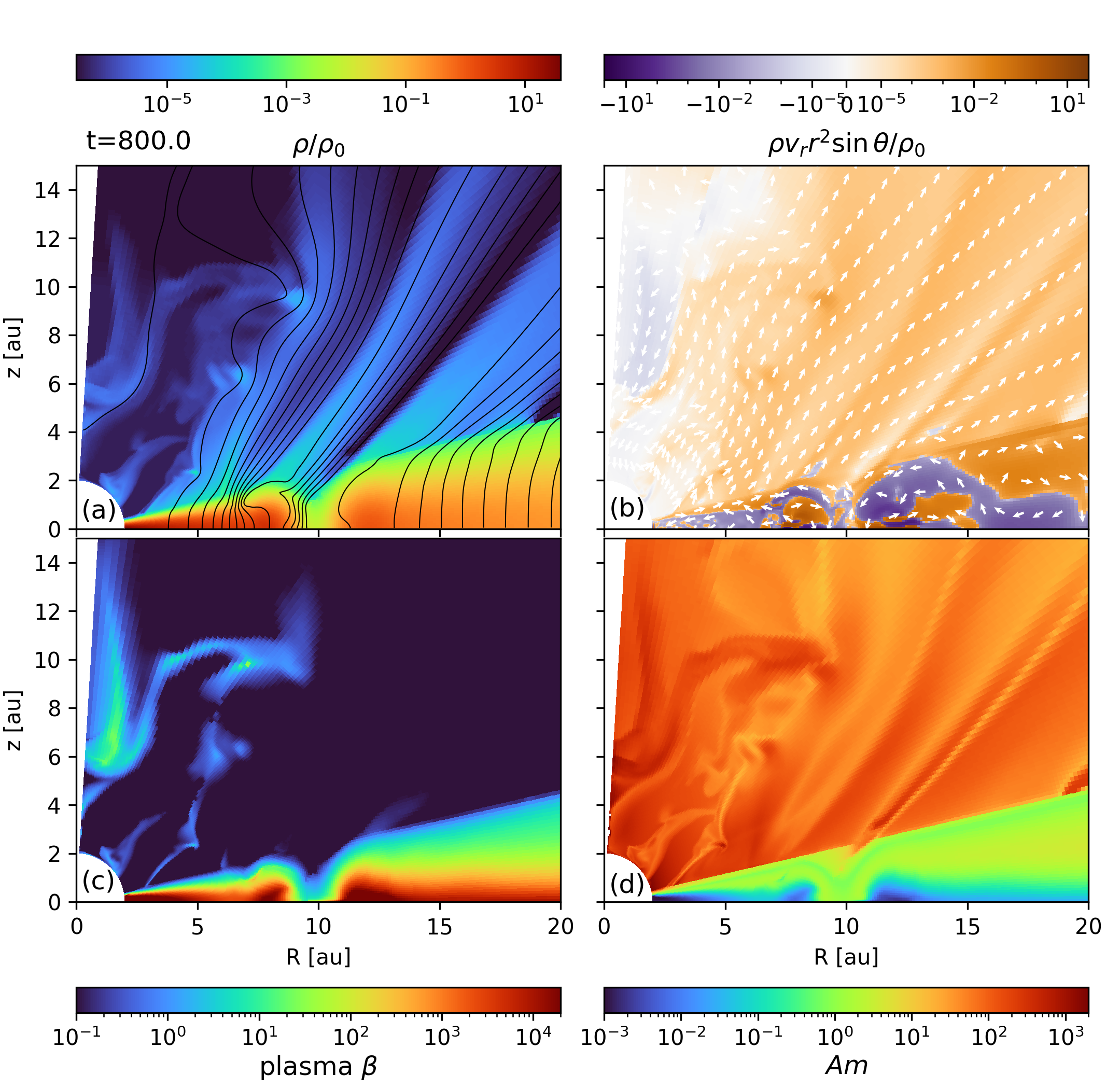}
    \caption{Properties of the fiducial model at a representative time of 800~years on the meridian plane. {\bf Panel (a)} shows the mass density $\rho$ scaled by a reference value $\rho_0$ (initial midplane density at R=10~au) and poloidal magnetic field lines (black); {\bf Panel (b)} the radial mass flux per unit area in code units, with the velocity unit vectors (white arrows) superposed; {\bf Panel (c)} the plasma-$\beta$; and {\bf Panel (d)} the Elsasser number $Am$. See the supplementary online material for an animated version of this figure. }
    \label{fig:overview}
\end{figure*}

In Fig.~\ref{fig:overview}, we plot several quantities on the meridian plane at a representative time of 800~years.
Several broad features are worth noting. First, as expected, a clear gap is created near the planet at 10~au, as seen in the density map shown in panel (a). Second, the wind is driven off the disk's surface along the initially imposed large-scale poloidal magnetic field lines, with a generally positive radial mass flux, as shown in panel (b). The flow pattern inside the disk is more complex, however, with patches of inflow (with a blue color in panel [b]) next to patches of outward expansion (colored brown). Third, the poloidal field lines have a clear concentration in the low-density gap, as shown in panel (a). The concentration continues into the wind. The combination of flux concentration and low density in the gap means a large drop in the plasma-$\beta$ compared to the surrounding disk regions, as illustrated in panel (c). The low density enables a higher ionization level, leading to a larger Elsasser number $Am$, as shown in panel (d). Indeed, the $Am$ value approaches or even exceeds unity, much higher than the surrounding disk regions, where $Am$ is well below unity. Therefore, the planet created a gap where the magnetic field is expected to be more dynamically significant and better coupled to the gas than its surroundings.

\begin{figure*}
    \centering
    \includegraphics[width=1.0\textwidth]{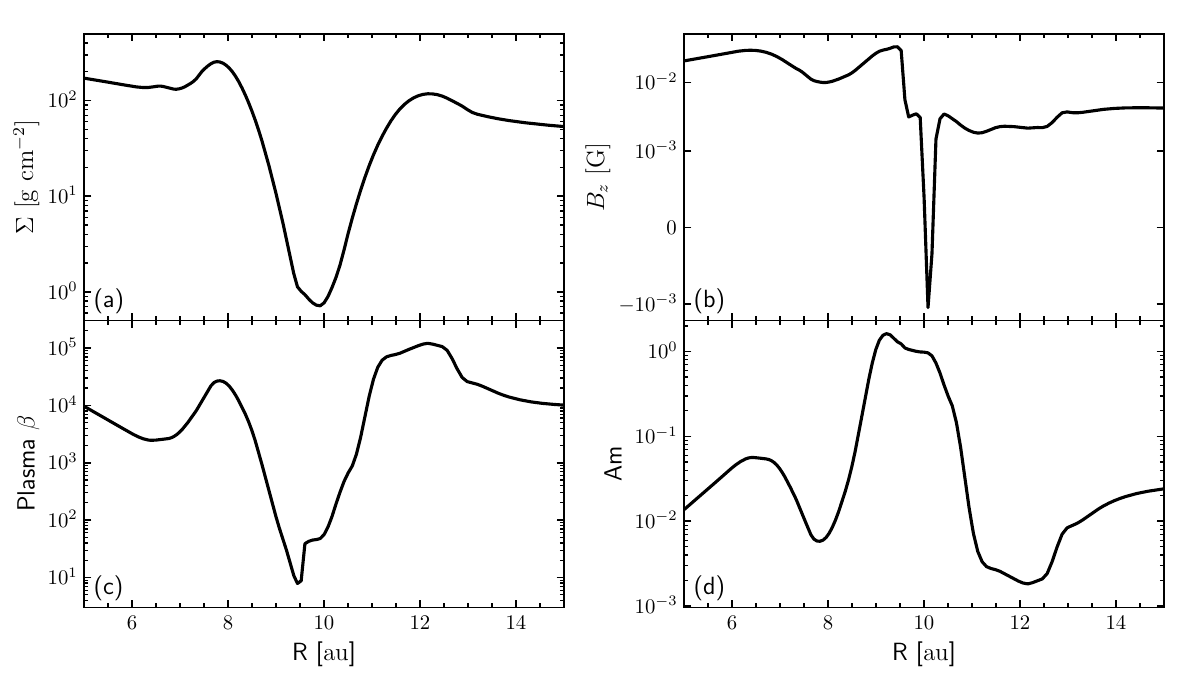}
    \caption{1D plots of the fiducial model at a representative time of 800~years. {\bf Panel (a)} shows the column density $\Sigma$; {\bf (b)} the vertical component of the magnetic field $B_{z,mid}$ (in Gauss) on the disk midplane; {\bf (c)} the plasma-$\beta$ on the disk midplane; and {\bf (d)} the Elsasser number $Am$ on the midplane.}
    \label{fig:faceon_view}
\end{figure*}

To illustrate the magnetically significant gap more quantitatively, we plot in Fig.~\ref{fig:faceon_view} the 1 D distributions of the column density $\Sigma$, the vertical component of the magnetic field $B_{z, mid}$ on the midplane, and the plasma-$\beta$ and Elsasser number $Am$ at the same location. The figure clearly shows the strong correlation between the column density (panel a) and the plasma-$\beta$ (c) and their anti-correlation with the field strength (b) and the Elsasser number (d) that were already apparent in the earlier meridional plot (Fig.~\ref{fig:overview}) It also reveals a more subtle difference between the dense ring immediately outside the gap (around $r\sim 12.5$~au) and that inside it ($r\sim 8$~au). The former has a much weaker midplane field strength (panel b) and a much lower Elsasser number (panel d) than the latter. The lower Elsasser number may appear surprising since the outer ring is less dense compared to the inner one and thus should be better ionized by high energy radiation and cosmic rays. However, the Elsasser number depends on not only the charge densities but also the magnetic field strength. In particular, a weaker field reduces the Hall parameter (see equation~\ref{eq:beta_j}), making it harder for the charged particles (especially the more massive charged grains) to couple to the field lines. We suspect that a feedback loop may be operating in the outer ring where a lower Elsasser number leads to faster diffusion of the magnetic field from the region, lowering the field strength, which, in turn, leads to a weaker field-matter coupling and, thus, an even lower Elsasser number.

\subsection{Origins of Gap Magnetic Flux Concentration}
\label{sec:FluxCon}
\begin{figure}
    \centering
    \includegraphics[width=0.45\textwidth]{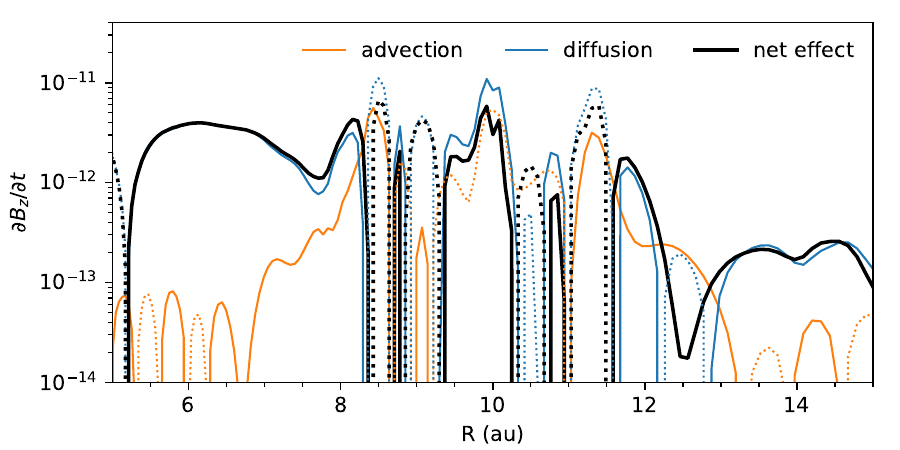}
    \caption{The blue lines correspond to the advection effect, while the orange lines represent magnetic diffusion. The black lines depict the net effect ($\partial B_z/\partial t$) of these two phenomena. Solid lines signify positive values, and dotted lines indicate negative values. The diffusion and advection terms are directly from Eq.~\ref{eq:dbzdt}.}
    \label{fig:diffbz}
\end{figure}
To help understand the magnetic concentration in the gap, we examine the time evolution of the vertical field component $B_z$ on the midplane. Under the assumption $B_R=B_\phi=0$ (where $R$ is the cylindrical radius), which is strictly true for an axisymmetric disk with a mirror symmetry for the upper and lower hemispheres, as assumed here), the induction equation (Eq.\ref{eq:induction}) can be reduced to:
\begin{eqnarray}
\frac{\partial {B_z}}{\partial t}=-\frac{1}{R}\frac{\partial \left(Rv_RB_z\right)}{\partial R} +
\frac{1}{R}\frac{\partial \left(R\left(\eta_O+\eta_A\right)\partial_RB_z\right)}{\partial R}.
\label{eq:dbzdt}
\end{eqnarray}
The right side of the equation consists of two terms: the first term represents advection, and the second term accounts for magnetic diffusion. We have plotted these terms separately and their sum in Figure~\ref{fig:diffbz}. To illustrate the trend of magnetic field evolution, the data is the average of 50 snapshots at relatively early times, from t=100 to t=150 years. The tendency for the vertical field strength to increase in the gap around the planet is already clear at these early times, with the magnetic diffusion causing $B_z$ to increase with time faster than the advection causing $B_z$ to decrease with time. This analysis indicates that magnetic diffusion is the root cause of the magnetic flux concentration in the gap. Therefore, its proper treatment is needed for quantifying the degree of flux concentration.

\subsection{Gas Dynamics in and around the Gap}
\begin{figure*}
    \centering
    \includegraphics[width=\textwidth]{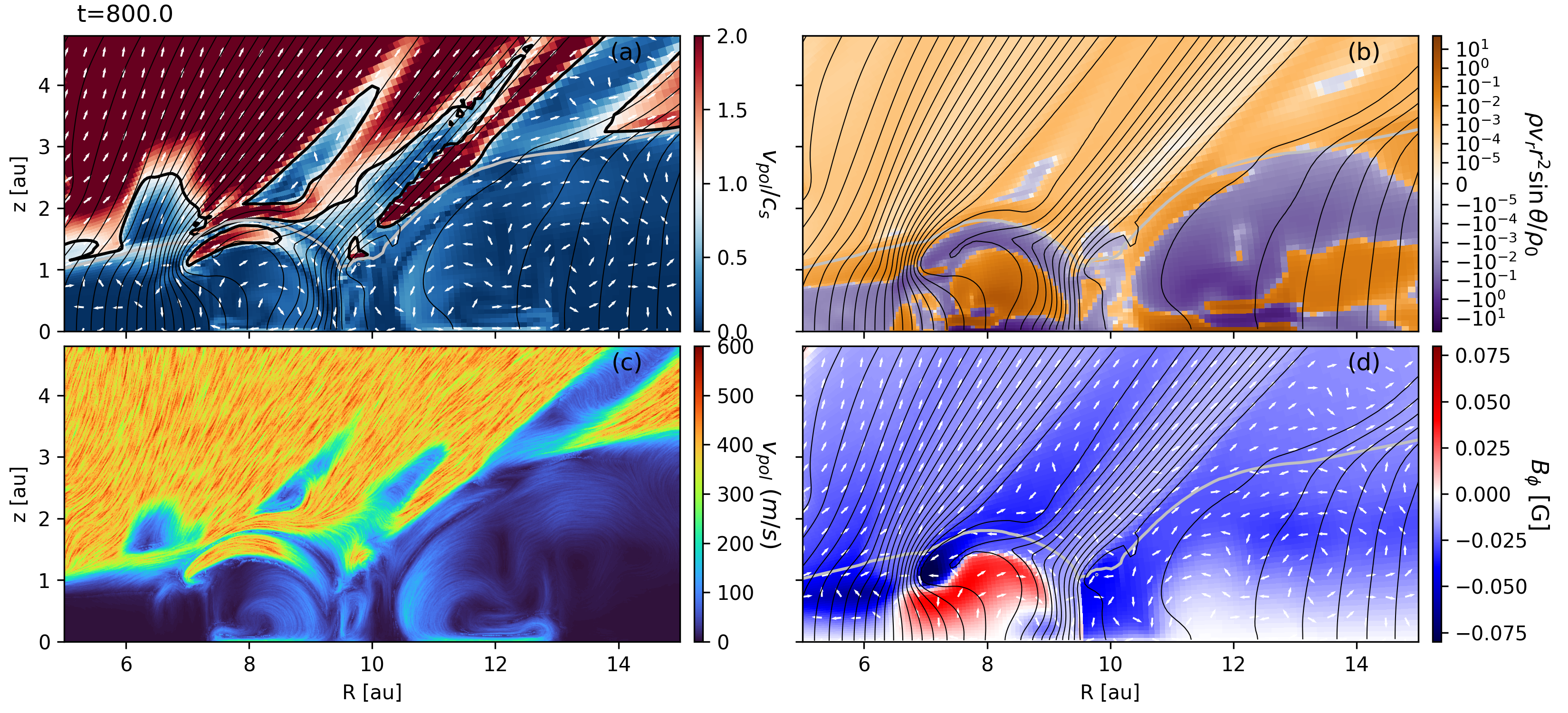}
  
    \caption{{\bf Panel (a)}: Gas poloidal velocity divided by the local sound speed, with arrows indicating the flow direction; {\bf (b)}: Effective radial mass flux (in code units), with positive indicating outward motion and the solid black lines showing magnetic field lines; {\bf (c)} Line integration contours (LIC) of the poloidal streamlines showing the magnetically-mediated meridional gas circulation near the planet, the area with a velocity above 600 m/s is not saturated by color due to the properties of LIC; {\bf (d)} Toroidal magnetic field strength (color map, in Gauss) with the field lines (black lines), velocity field (white vectors), and an isodensity contour (thick gray line) to help mark the low-density gap.
    }
    \label{fig:gapvec_fid}
\end{figure*}
\begin{figure*}
    \centering
    \includegraphics[width=1.0\textwidth]{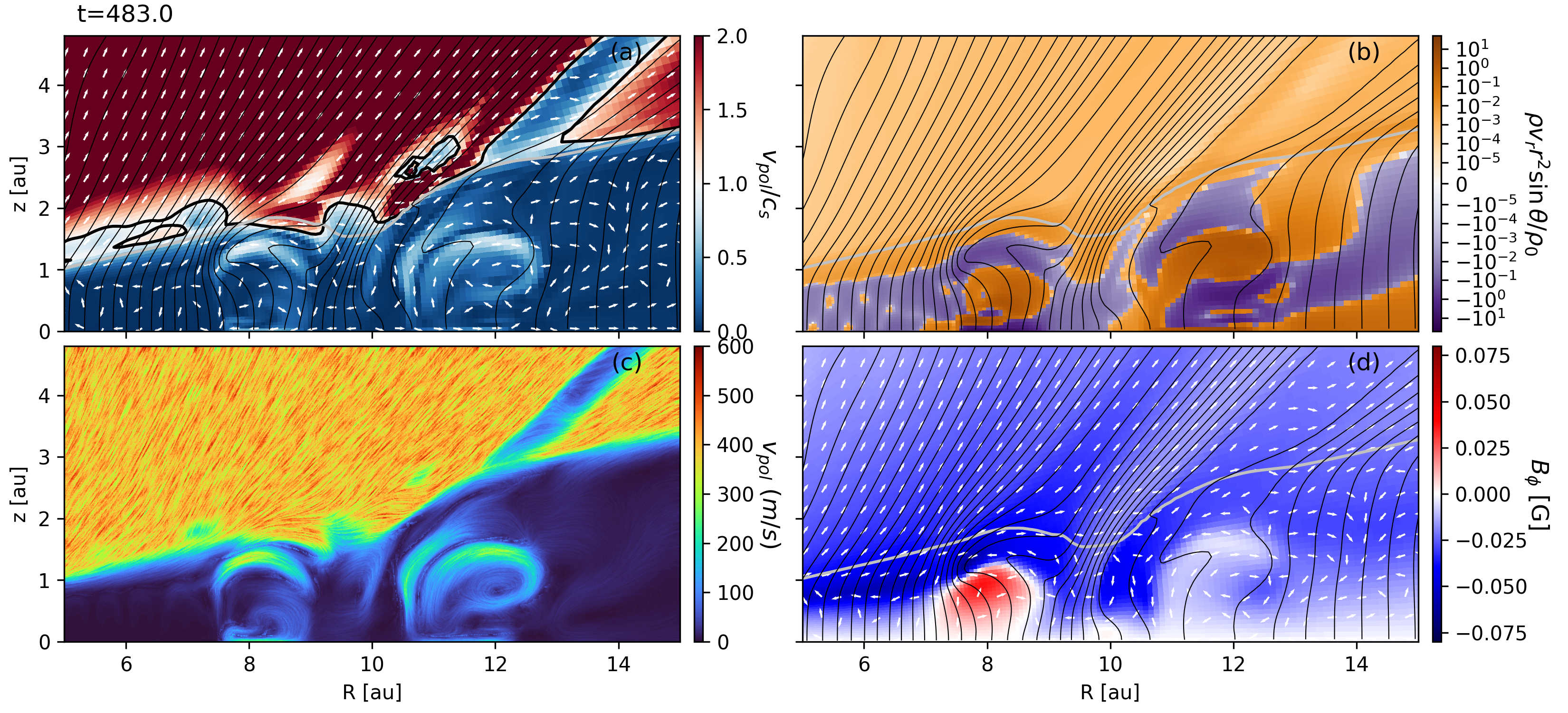}
  
    \caption{Same as Fig.~\ref{fig:gapvec_fid} but at an earlier time of 483~years, showing prominent magnetically-driven fast (sonic) meridional gas circulations near both inner and outer edges of the planet-opened gap.  }

    \label{fig:gapvec_fid_f400}
\end{figure*}

\begin{figure*}
    \centering
    \includegraphics[width=1.0\textwidth]{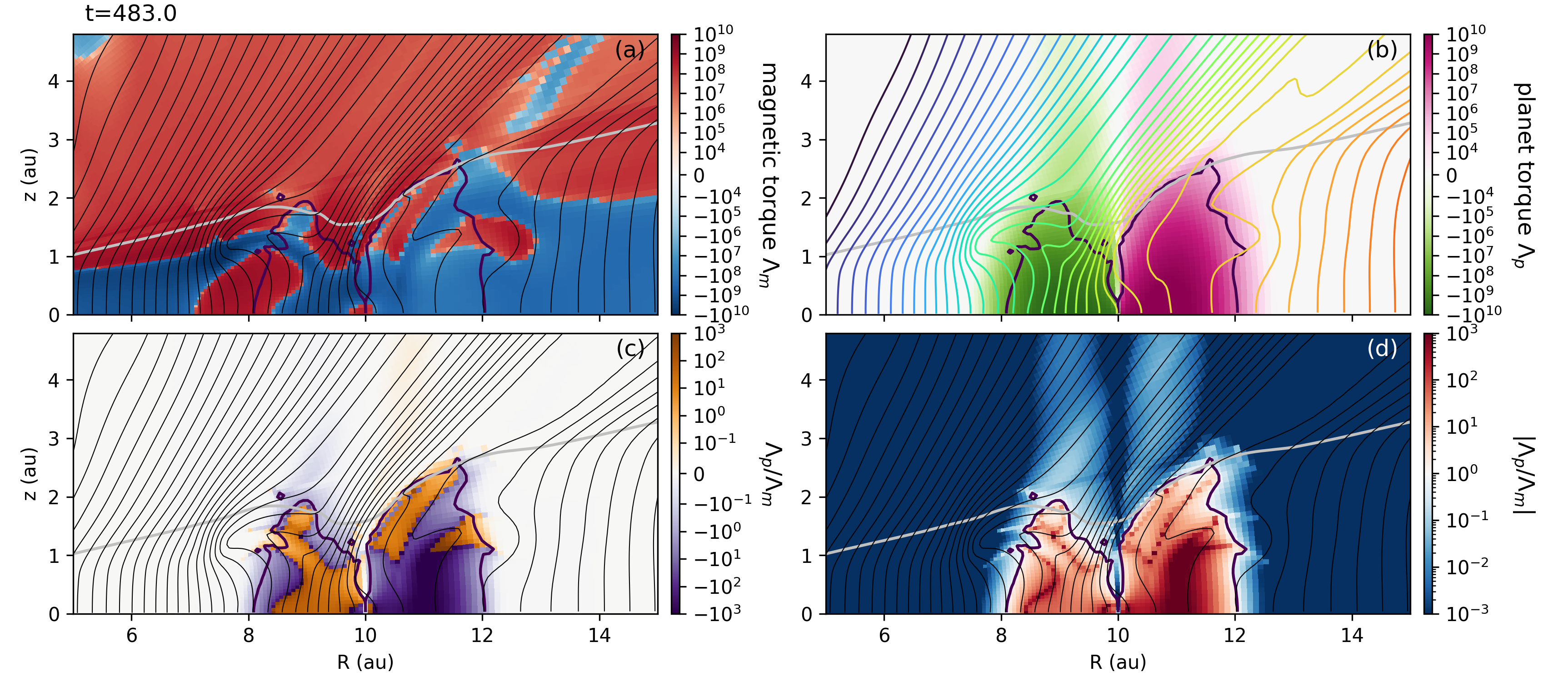}
    \caption{Torque (in code units) distribution comparison at t=483. {\bf Panel (a)} and {\bf (b)} are magnetic torque $\Lambda_m$ and planet torque $\Lambda_p$ plotted in the same symmetric log scale; {\bf (c)} shows the ratio of planet torque over magnetic torque, and {\bf (d)} illustrated the absolute value of the same ratio. The field lines are black solid lines in all panels but {\bf (b)}, where they are color-coded to indicate the magnetic flux within (to the inner boundary) each line and have an offset to the black lines in the other panels. See the supplementary material online for an animation of the figure.}
    \label{fig:torquecomp}
\end{figure*}

To illustrate the gas dynamics in and around the planet-opened gap, we plot in Fig.~\ref{fig:gapvec_fid} the poloidal flow speed (color map) and flow directions (arrows) in a meridional plane in a zoomed-in region around the gap. Part of the flow pattern is similar to the non-magnetized model; the gas is repelled away from the planet near the midplane and circulates back towards the planet at higher altitudes. However, there are several significant differences. First, near the inner edge of the gap, there is a fast (transonic) surface accretion stream. It originates from about one scale height above the midplane, with an initial speed of around a quarter of the sound speed. The stream accelerates as it rises to a higher altitude and then slides inwards along the disk surface, reaching a supersonic speed (Figure~\ref{fig:gapvec_fid}a).
Interestingly, the accreting stream (with a negative radial mass flux in panel b) is located close to a sharp kink in the poloidal magnetic field lines (panel b). The spatial correlation is not a coincidence; it is expected because the material near the tip of the kink that opens outward rotates faster than the material along the same field line but at larger radii, causing it to lose angular momentum and fall inward. Part of the angular momentum is transported out of the disk along field lines to the low-density wind. The remaining part is transported along field lines to the denser disk region below the accretion stream, causing it to expand. This magnetically mediated redistribution of angular momentum \footnote{The phenomenon is essentially the magnetically driven surface ``avalanche flow" described in \citet{1996ApJ...461..115M} \citep[see also][and references therein]{2017MNRAS.468.3850S}. It is closely related to the well-known channel flows in early simulations of the magneto-rotational instability \citep[e.g.,][]{1992ApJ...400..595H}.} can be seen most clearly in Panel (d) of the Figure, where we plot the toroidal magnetic field in the color map. Note the negative $B_\phi$ in the accretion stream, indicating that it is magnetically braked.
In contrast, the region below the stream has a positive $B_\phi$, indicating that the magnetic field is trying to spin up the material, causing it to expand outward. The thermal pressure gradient can aid this outward expansion. The positive magnetic torque and pressure gradient combine to overcome the persistent negative planetary torque in the region that tends to remove its angular momentum and drive it to accrete. Similarly, the surface accretion stream may be helped by the negative planetary torque, which aids the negative magnetic torque in removing its angular momentum and thus drives the stream to accrete faster than it would otherwise. The net effect of the interplay between the magnetic and planetary torques is the production of fast meridional motions above the circulation cell driven directly by the planet close to the midplane (see panel [c]).

The surface accretion stream and associated meridional circulations are persistent features of our simulations (see the animated version of Figure~\ref{fig:gapvec_fid}). It is not limited to the inner edge of the gap; it often occurs near the outer edge as well, especially at earlier times. For example, we show in Fig.~\ref{fig:gapvec_fid_f400} the same plots as in Fig.~\ref{fig:gapvec_fid} but for an earlier time of $t=483$~years. Note the double-kinked, ``S-shaped" poloidal field line \citep[similar field geometry is also reported in][]{2022A&A...667A..17M,2019ApJ...885...36H} near the radius of 11 au outside the planet, with a fast, nearly sonic infall around the upper field line kink that opens outward and a fast expansion around the lower kink that opens inward. As discussed earlier, this is the expected behavior of magnetically mediated angular momentum redistribution, which can drive meridional gas circulation up to the sound speed near both the inner and outer edges of the planet-opened gap (see panel [a]). 

To illustrate the role of magnetic fields on the gas meridional circulation further, we plot in Fig.~\ref{fig:torquecomp} the distribution of the magnetic torque (panel a) at the same time shown in Fig.~\ref{fig:gapvec_fid_f400}. As expected, regions near the gap where the magnetic field is braking the disk rotation (with a negative torque) tend to infall, while those with a positive magnetic torque tend to expand (compare Fig.~\ref{fig:torquecomp}a with Fig.~\ref{fig:gapvec_fid_f400}b). The correlation is not perfect, though, because of the presence of a thermal pressure gradient and a persistent planetary torque (Fig.~\ref{fig:torquecomp}b), which dominates the magnetic torque near the gap (inside the thick black contours of Fig.~\ref{fig:torquecomp}c), particularly in the relatively high-density regions close to the midplane inside the gap and near the inner and outer edges of the gap and the surrounding rings. However, even in the regions with higher planetary torque, the magnetic torque remains significant, with a value typically smaller by a factor of only a few compared to the planetary torque (see panel d). Interestingly, even in such regions, there are fast (sonic) flows toward the planet despite the tendency for the planetary torque to push them away from the planet. As mentioned earlier, these flows are likely helped by strong pressure gradients near the gap edges, as in the hydro case. 

It is worth noting that, at the (relatively early) time shown in Fig.~\ref{fig:torquecomp}, the magnetic torque dominates the planetary torque in a roughly ``V-shaped" region directly above the planet (see panel c) because of a relatively low density and strong magnetic field in this part of the gap (see Fig.~\ref{fig:torquecomp}). It is a region completely dominated by the magnetic field, with a plasma-$\beta$ below unity and an Elsasser number above unity. As the density in the gap continues to drop at later times, this magnetically-dominated region extends all the way to the disk midplane, forming a roughly ``U-shaped" region that centers not on the planet but to a location interior to it, where most of the poloidal magnetic flux in the gap is concentrated (see, e.g., the time $t=1000$~years in the animated version of Fig.~\ref{fig:torquecomp} online). 

The poloidal magnetic flux concentration in the gap results from the non-ideal MHD effects (ambipolar diffusion and Ohmic dissipation) included in our simulation. The reason is that the flux-to-mass ratio is conserved in the ideal MHD limit, with a lower flux expected in the strongly mass-depleted gap. This argument is consistent with our analysis in \S\ref{sec:FluxCon}, although a complete understanding of exactly how the needed redistribution of the magnetic flux relative to the mass occurs is still lacking. From the evolution of the poloidal magnetic field lines, we observed that some field lines are dragged into the gap from larger radii by accretion flows and get stuck. Why the dragged-in field lines stay in the gap is unclear, but it is consistent with other non-ideal MHD disk simulations that include ambipolar diffusion where the poloidal field strength strongly anti-correlates with the surface density \citep[e.g.,][]{2018MNRAS.477.1239S,2020A&A...639A..95R,2021MNRAS.507.1106C,2022MNRAS.516.2006H, 2023ApJ...946....5A}.

\subsection{Magnetic Diffusivity-Gas Density Relation}
\label{sub:diffusivity}
\begin{figure}
    \centering
    \includegraphics[width=0.45\textwidth]{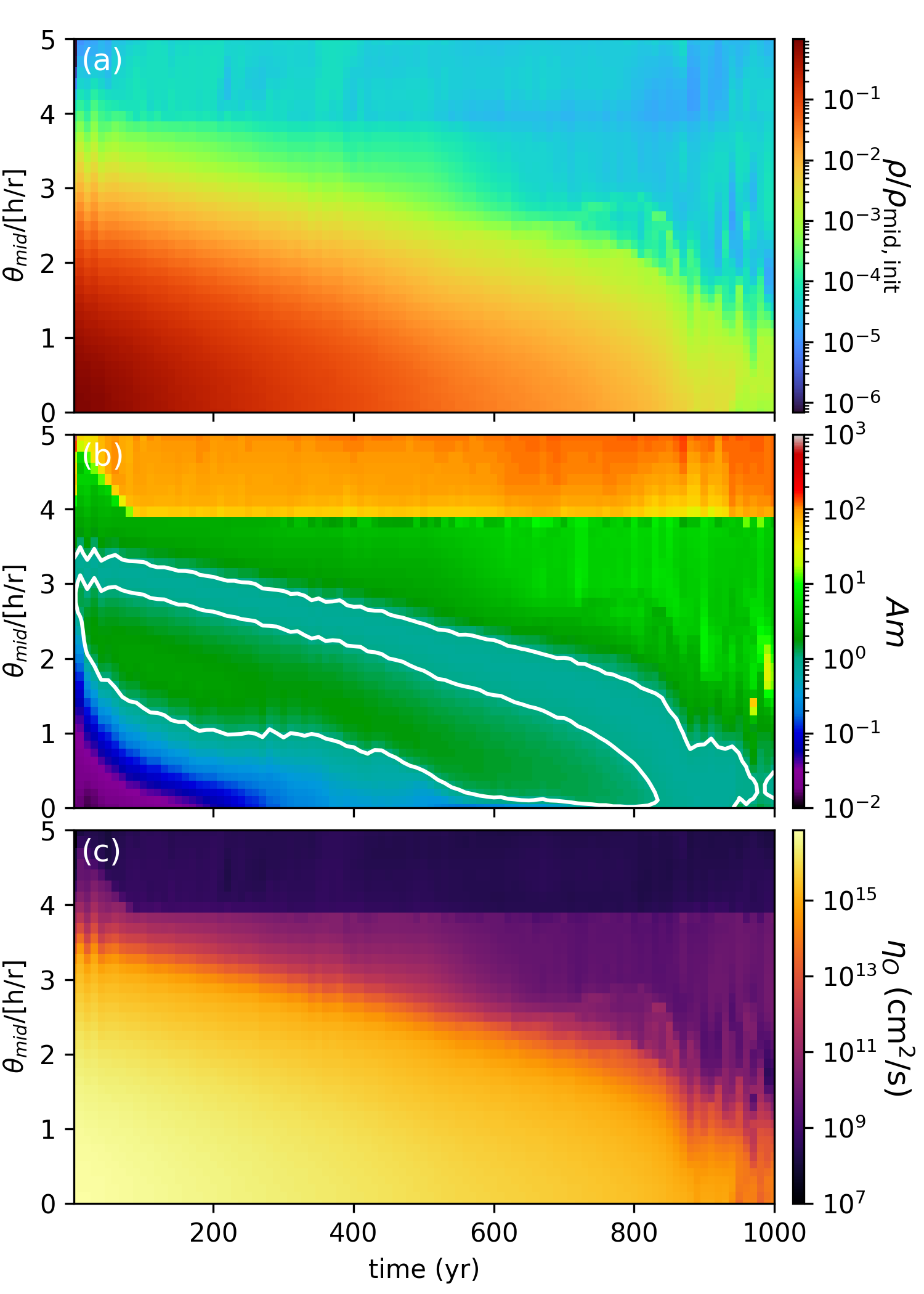}
    \caption{Time evolution of the gas density (panel a), $Am$ (panel b), and $\eta_O$ (panel c) at different heights at the radius $r=10$~au. The white contours in panel (b) indicate $Am=1$.}
    \label{fig:rhoamtime}
\end{figure}
A major advantage of our 2D (axisymmetric) non-ideal MHD simulation enabled by the prescribed planetary torque is that it allows us to evolve the disk thermochemistry (including ionization) self-consistently. In contrast, most existing disk simulations that include nonideal MHD processes either use spatially dependent only diffusion profiles \citep[e.g.,][]{2017ApJ...836...46B,2021MNRAS.507.1106C} or simple power-laws that depend only on the local density \citep[e.g.,][]{2018MNRAS.477.1239S,2022MNRAS.516.2006H} or column density \citep[e.g.,][]{2020A&A...639A..95R}. In this section, we approximate the numerically obtained magnetic diffusivities as a function of the local density using analytic fitting formulae. Such formulae will be useful for future 3D studies where it is too computationally expensive to compute the thermochemistry self-consistently in each cell and at each time step. 

 To illustrate the dependence of the magnetic diffusivities on the local density, we plot in Figure~\ref{fig:rhoamtime} the time evolution (up to 1000 years) of the disk mass density $\rho$, the Elsasser number $Am$, and the Ohmic diffusivity $\eta_O$ as a function of the polar angle $\theta_{mid}$ away from the mid-plane (in units of the ratio of the scale height $h$ over the radius $r$) at the radius of $r=10$~au (where the planet is located). The sharp change of $Am$ and $\eta_O$ at $\sim 4$~disk scale heights corresponds to the height of UV penetration. The Ohmic diffusivity shown in Panel (c) appears to correlate with the gas density shown in Panel (a), with the diffusivity increasing nearly monotonically with the density. The change of $Am$ with the density appears more complex. While $Am$ generally anti-correlates with $\rho$, there is an $Am$ ``valley" just below the disk surface, starting from a height of $\sim$3h and reaching below $\sim$2h after 800 years. This trend resembles the power-law prescription proposed by \citet{2018MNRAS.477.1239S} where the $Am$ inside the disk decreases with a decreasing density as a result of balancing the recombination of the dominant ions and electrons with the cosmic ray ionization.

\begin{figure}
    \centering
    \includegraphics[width=0.45\textwidth]{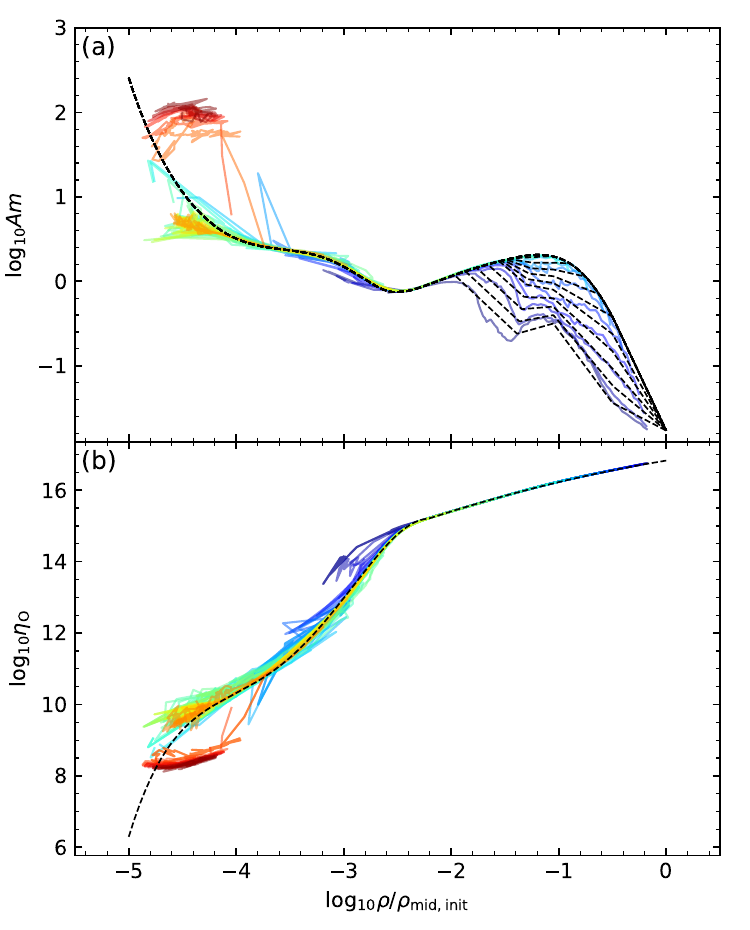}
    \caption{Panel (a) shows the relation between density and $Am$, while Panel (b) shows the relation between density and $\eta_O$. The measurement is taken from the midplane (blue lines) to $\sim 5$~scale heights (red lines) at an increment of 0.25 scale height.}
    \label{fig:etaamrho}
\end{figure}

To better understand the magnetic diffusion-density relation, we plot the $Am-\rho$ and $\eta_O-\rho$ diagrams in Figure~\ref{fig:etaamrho}. Each track in the diagrams corresponds to the time evolution of $Am$ and $\rho$ or $\eta_O$ and $\rho$ at a given height at $r=10$~au, with the height denoted by the track's color (the bluest corresponds to the midplane and the reddest marks $\sim$ 5.5h, with an increment of 0.25 scale height). The Ohmic diffusivity in Panel (b) has a simple dependence on the gas density: almost a single monotonic track. It means $\eta_O$ barely depends on its vertical location. We used two polynomials,  $\mathcal{P}_{\rm O,1}$ and $\mathcal{P}_{\rm O,2}$, to fit the Ohmic-density relation in the log-log space, with the coefficients $k_i$ of the polynomials ($\mathcal{P}=\sum_{i=0}^{n}k_ix^i$) given in Table \ref{table:kn}).

The $Am-\rho$ diagram is a different story. In Panel (a) of Figure~\ref{fig:etaamrho}, the tracks representing different heights from the midplane only start to converge into a single track when the density drops below the initial midplane density by about two orders of magnitude. The multiple ion species and charged grains acting like ``heavy ions" give the ambipolar diffusion much more complicated behaviors than the Ohmic diffusivity. To have a better-performing look-up table of $Am-\rho$, we ignored the $Am$ evolution in the first 100 yrs. The fitting result is given in Table \ref{table:LUT} and shown in Figure \ref{fig:etaamrho} as dashed lines. This table\footnote{The table is available in both Python and C scripts at \url{https://github.com/astroxhu/diffusion-table}} could be used for future 3D protoplanetary disk simulations covering a high dynamic gas density range, especially near the midplane. Note the look-up table is suitable for the outer part of the disk that is relatively well shielded from the central ionization source. Care must be exercised when applying it to the innermost part of the disk or a transitional disk with a large inner hole, as the shielding of the central ionization source may be significantly different from that envisioned in the model.

\begin{table*}
  \caption{Am-$\rho$ look-up table (\S\ref{sub:diffusivity}) 
    \label{table:LUT}
  }
  \centering
  \scriptsize
\begin{tabular}{l c r}
\hline
\hline
  $\log_{10}{\rho^*}$ & $\theta^*$ & $\log_{10}{Am}$\\
  \hline
  -0.5 to 0. & 0 to 4. & $(-2.04\theta^*_1 - 0.656)\times\log_{10}{\rho^*} -1.768$\\

  -1.05 to -0.5 & 0 to 0.8 & $(0.945\theta^* - 1.709)\times\log_{10}{\rho^*} +1.493\theta^* - 2.295$\\
  -1.05 to -0.5 & 0.8 to 4 &
  $(1.127\theta^*_1 - 0.953)\times\log_{10}{\rho^*}+\
        1.584\theta^*_1 - 1.1$\\
  $\mathcal{F}(\log_{10}{\rho^*},\log_{10}{Am})$ to -1.05 & 0 to 0.8 & $(0.121 - 0.227\theta^*_1)/(0.338 - 0.11\theta^*_1)\times\log_{10}{\rho^*}+\
        0.5\theta^*_1 + 1.05\times(0.121 - 0.227\theta^*_1)/(0.338 - 0.11\theta^*_1) - 0.5$\\
  $\mathcal{F}(\log_{10}{\rho^*},\log_{10}{Am})$ to -1.05& 0 to 0.8 &
  $(0.0756\theta^*_2 - 0.0606)/(0.25 - 0.0566\theta^*_2)\times\log_{10}{\rho^*}+\
        0.4\theta^*_2 - 0.1 + 1.05\times(0.0756\theta^*_2 - 0.0606)/(0.25 - 0.0566\theta^*_2)$\\
    -2.5 to $\mathcal{F}(\log_{10}{\rho^*},\log_{10}{Am})$ & 0 to 0.8 & $(0.537\theta^* - 0.702)/(0.587 - 0.421\theta^*)\times\log_{10}{\rho^*}+\
          0.727\theta^* - 0.621$\\ & & $ - (0.11\theta^* - 1.388)*(0.537\theta^* - 0.702)/(0.587 - 0.421\theta^*)$ \\
     -2.5 to $\mathcal{F}(\log_{10}{\rho^*},\log_{10}{Am})$ & 0.8 to 4 &  $(0.2864\theta^*_2 - 0.2724)/(0.2502 - 0.0496\theta^*_2)\times\log_{10}{\rho^*}+\
        0.3244\theta^*_2 - 0.0394 $\\ & &$-(0.0566\theta^*_2 - 1.3)\times(0.2864\theta^*_2 - 0.2724)/(0.2502 - 0.0496\theta^*_2)$\\    
 -2.5 to -0.5 & 0 to 4 & Upper Limit : $\mathcal{P}_{\rm AD,1}$, see Table~\ref{table:kn}\\
 $\log_{10}{\rho^*_{atm}}$ to -2.5 & 0 to 4 & $\mathcal{P}_{\rm AD,2}$, see Table~\ref{table:kn}\\
  
  \hline
  \hline
  \end{tabular}
  
  \smallskip
\begin{tabular}{p{0.95\linewidth}}
Here $\rho^*=\rho/\rho_{\rm mid,init}$, $\theta^*=\theta/(h/r)$, $\theta^*_1=\min{(\theta^*,1)}$, $\theta^*_2=\min{(\theta^*,1.95)}$. $\mathcal{F}$ is the equation of a straight line $6.61\log_{10}{\rho^*}-\log_{10}{Am}+8.55=0$.
\end{tabular}
\end{table*}

\begin{table}
\caption{Coefficients for polynomial fits of AD and Ohmic look-up table (\S\ref{sub:diffusivity}) 
    \label{table:kn}
  }
    \centering
    \begin{tabular}{ccccc}
    \hline
    \hline
    $k_i$ & $\mathcal{P}_{\rm AD,1}$ &$\mathcal{P}_{\rm AD,2}$ &$\mathcal{P}_{\rm O,1}$ & $\mathcal{P}_{\rm O,2}$ \\
    $k_6$&0.26429362&0.134528857&&\\
    $k_5$&2.308469&3.15915033&&\\
    $k_4$&7.81413499&30.6754303&&-0.82772958\\
    $k_3$&12.39250353& 157.055386&&-11.48813613\\
    $k_2$&7.90442493&445.96139&-0.10996372&-58.24798054\\
    $k_1$&-0.52179895&664.2113&0.49186494&-124.86627111\\
    $k_0$&-1.52831614&404.541785&16.82894602&-80.52351483\\
         \hline
    \end{tabular}
\end{table}

\section{Discussion and Conclusions}
\label{sec:discussion}

\begin{figure*}
    \centering
    \includegraphics[width=\textwidth]{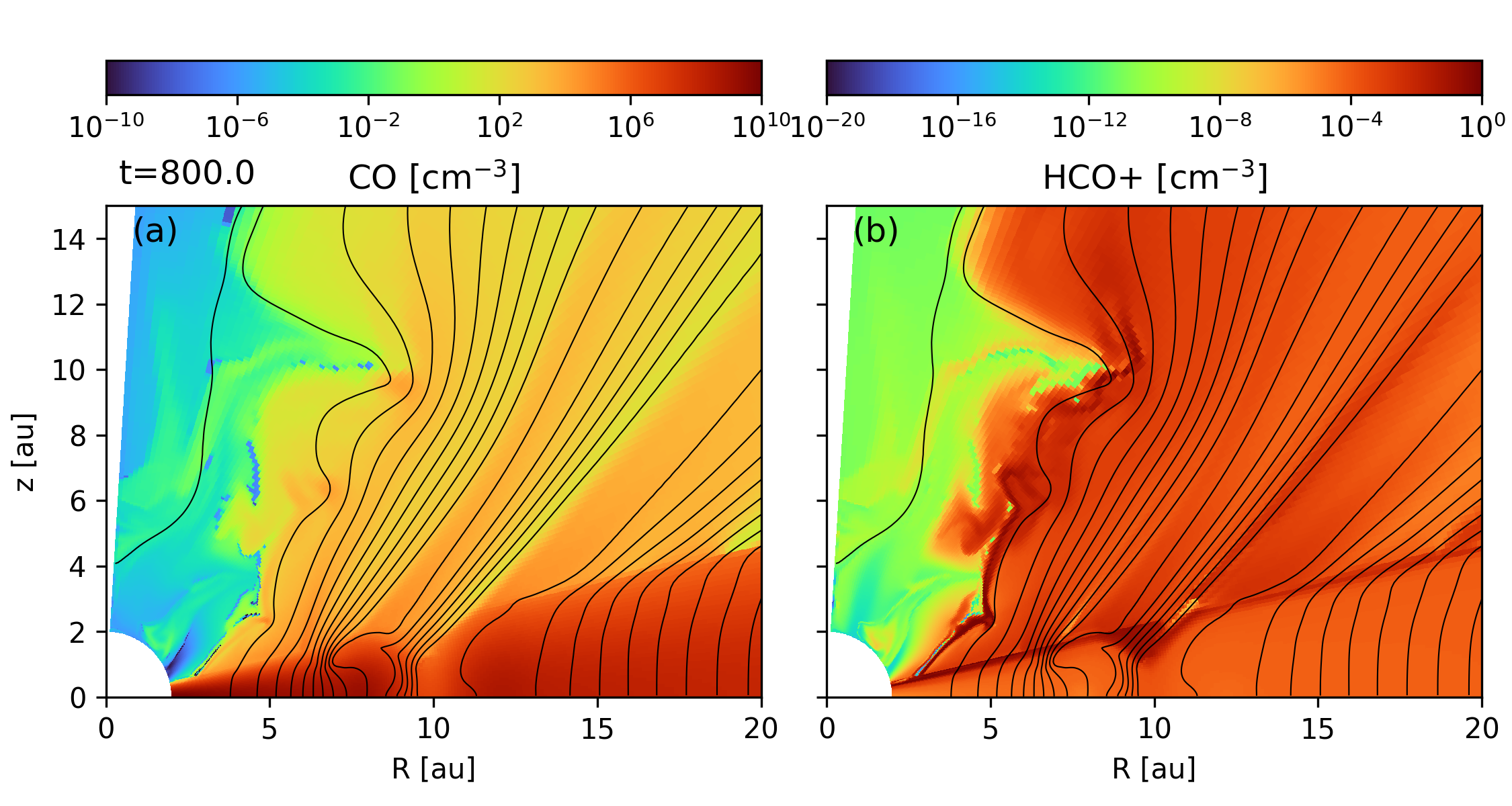}
    \caption{Distributions of the number densities of CO (Panel a) and HCO$^+$ (Panel b) at a representative time $t=800~$years, showing enhanced HCO$^+$ in the planet-opened gap where the CO density is reduced along with the gas.}
    \label{fig:COHCO}
\end{figure*}

Our consistent thermochemical modeling of the planet's gap opening allows us to determine various chemical species' spatial distributions and evolution in the disk and its surrounding environment. Of the modeled species, the most directly relevant to observations are CO and HCO$^+$. Their number densities are shown in Fig.~\ref{fig:COHCO}. As expected, the CO number density is lower in the gap than in the denser surrounding regions because of gas depletion. However, the HCO$^+$ number density is higher in the gap than in the surrounding regions despite a lower gas density, indicating a much higher fractional abundance. The enhanced HCO$^+$ abundance is consistent with the ALMA observations of AS 209, one of the protoplanetary disks with prominent gaps potentially opened by planets, which show that HCO$^+$ has an unusually high abundance in the radial range between $\sim 50$ and $\sim 150$~au where the CO column density shows a depression \citep{2021ApJS..257...13A}. It is also in agreement with the greatly enhanced DCO$^+$ in the same region \citep[see their Fig.~4]{2019ApJ...871..107F}. 

\begin{figure}
    \centering
    \includegraphics[width=0.45\textwidth]{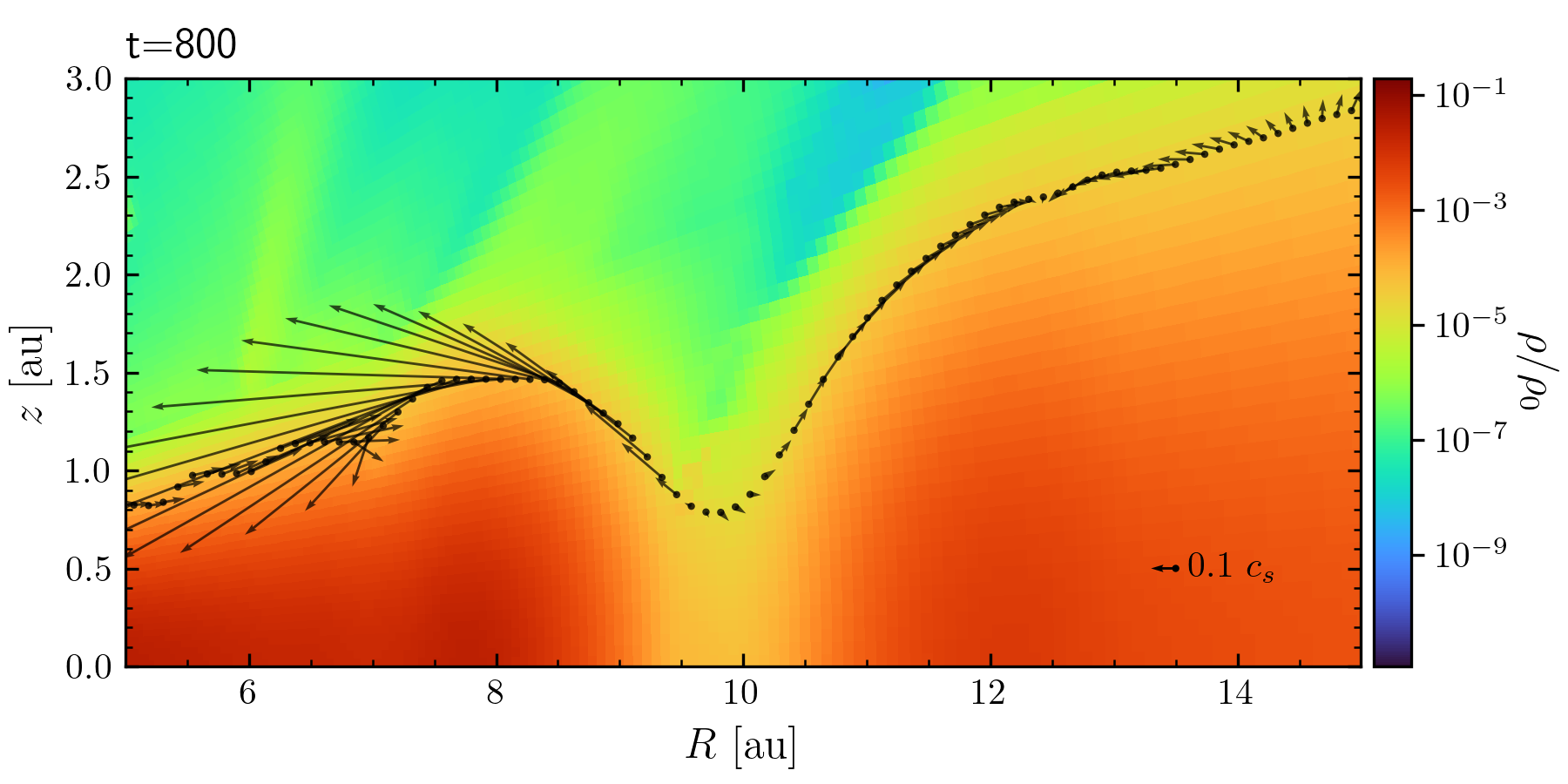}
    \caption{Gas velocity structure presented the same general style as in Fig.~2 of \citet{2019Natur.574..378T} to facilitate comparison with observations. Gas velocity vectors in the meridional plane at the surface, with a vector for 0.1~$c_s$ shown for reference. The color map in the background is the gas density.}
    \label{fig:obsvec}
\end{figure}

To better connect with observations, we follow \citet{2019Natur.574..378T} and plot the gas velocity structure in $R-z$ plane in Figure~\ref{fig:obsvec}, where $R$ is the cylindrical radius and $z$ the height above the midplane. For illustration purposes, we picked a surface that has a constant column density above it (from the top of the simulation domain). 
This figure is to be compared with the Fig.~4 of \citet{2023arXiv230403665G} for the AS 209 disk. Since our simulation was not explicitly designed for the AS 209 system, a detailed match is not to be expected. Nevertheless, we find velocity vectors pointing upward and inward at the gap's inner edge, which resembles the nearly sonic meridional flow observed in the $^{12}{\rm CO}$-emitting region of the AS 209 disk. However, the current model does not capture the outflow in the low-density gap region probed by $^{13}$CO, possibly because the adopted planetary torque profile makes the gap region too evacuated (and thus too strongly magnetically dominated) to launch an outflow efficiently. 

Our work is most directly comparable to \citet{2023ApJ...946....5A}, who carried out 3D global simulations of gap opening by planets in non-ideal MHD disks with a spatially constant ambipolar Elsasser number (either 1 or 3) inside the disk. They find a concentration of the poloidal magnetic flux in the planet-opened gap, which they believe is likely associated with the spiral density shocks at the surface of their simulated disks. We also find a concentration of poloidal magnetic flux in the planet-opened gap, despite a different (and more self-consistent) treatment of the ambipolar diffusivity, indicating that it is a rather general phenomenon. In particular, it does not depend on spiral density shocks, which are absent in our 2D (axisymmetric) simulation. We believe it is caused by the magnetic flux diffusing into the gap and being trapped because of enhanced local magnetic coupling to the lower-density gas. However, the details of the process remain to be determined.  

In summary, we carried out 2D (axisymmetric) simulations of gap opening by a planet embedded in a wind-launching non-ideal MHD disk with a prescribed planetary torque but self-consistent thermochemistry. Our main conclusions are as follows:

1. There is a strong concentration of poloidal magnetic flux in the planet-opened gap, where the magnetic field is much better coupled to the gas than in the denser regions surrounding the gap and where the magnetic pressure becomes comparable to, or even larger than, the gas pressure. Combined with similar results recently obtained by \citet{2023ApJ...946....5A} under very different conditions, our finding indicates that the flux concentration and the resulting magnetic domination of the gas dynamics are robust features of the planet-opened gaps in non-ideal MHD disks. 

2. Magnetic fields are also dynamically important in the denser regions surrounding the planet-opened gap. They drive fast infall and expansion motions up to the local sound speed through magnetic torque-induced angular momentum redistribution, thereby controlling the meridional gas circulation near the inner and outer edges of the gap. 

3. The plant-opened gap has a much higher abundance of molecular ion HCO$^+$ than its denser surrounding regions, consistent with high-resolution ALMA observations of the protoplanetary disk AS 209 with prominent rings and gaps. The magnetically-induced fast accretion stream near the disk surface at the inner edge of the gap is also consistent with sonic meridional flows probed by the AS 209 $^{12}$CO observations.

4. We numerically obtained fitting formulae for the ambipolar and Ohmic diffusivities as a function of the local density that can be used for future 3D simulations of planet gap-opening in non-ideal MHD disks where thermochemistry is too computationally expensive to evolve self-consistently with the magneto-hydrodynamics.

\section*{Acknowledgements}
The authors would like to thank the referee for the helpful report. X.H. acknowledges support from the University of Virginia through VICO (Virginia Initiative on Cosmic Origins) and NSF AST-1815784.
ZYL is supported in part by NASA 80NSSC20K0533 and NSF AST-1910106. Our simulations are made possible by an XSEDE allocation (AST200032). L.W. acknowledges the computation resources provided by the KIAA. Z.Z. acknowledges support from the National Science Foundation under CAREER Grant Number AST-1753168 and support from NASA award 80NSSC22K1413. Figures in
this paper were made with the help of \verb|Matplotlib| \citep{Hunter2007} and
\verb|NumPy| \citep{2020Natur.585..357H}. 

\section*{Data Availability}

The data from the simulations will be shared on reasonable request to the corresponding authors.

\bibliographystyle{mnras}
\bibliography{ref_mhd_dust}


\bsp	
\label{lastpage}
\end{document}

%% file: preamble.tex
\usepackage{float}
\usepackage{amsmath,amssymb}
\usepackage{natbib}    
\usepackage{hyperref} 
\usepackage{graphicx} 
\usepackage{color}
\usepackage[verbose]{placeins}
\usepackage{stfloats}
\usepackage{verbatim}    
\usepackage{enumitem}
\usepackage{natbib}
\usepackage{CJK}
\usepackage{multirow}
\usepackage{comment}
\usepackage{bm}
\usepackage{physics}
\usepackage{array}
\usepackage{orcidlink}  

\newcolumntype{L}[1]{>{\raggedright\arraybackslash}p{#1}}
\newcolumntype{C}[1]{>{\centering\arraybackslash}p{#1}}
\newcolumntype{R}[1]{>{\raggedleft\arraybackslash}p{#1}}

\newcommand{\cntextsc}[1]
{\begin{CJK*}{UTF8}{gbsn}#1\end{CJK*}}
\newcommand{\proptosim}{\mathrel{\vcenter{
 \offinterlineskip\halign{\hfil$##$\cr
 \propto\cr\noalign{\kern2pt}\sim\cr\noalign{\kern-2pt}}}}}

\newcommand{\mean}[1]{\langle #1\rangle}

\renewcommand{\min}{\mathrm{min}}

\newcommand{\au}{\mathrm{AU}}
\newcommand{\cm}{\mathrm{cm}}

\newcommand{\K}{\mathrm{K}}

\newcommand{\eV}{\mathrm{eV}}
\newcommand{\keV}{\mathrm{keV}}
\newcommand{\s}{\mathrm{s}}

\newcommand{\ang}{\ensuremath{\text{\AA}}}

\renewcommand{\O}{\mathrm{O}}     
\newcommand{\A}{\mathrm{A}}     
\renewcommand{\H}{\mathrm{H}}     
\renewcommand{\P}{\mathrm{P}}     

\newcommand{\sigperH}{$\sigma_\Gr/\H$}

\newcommand{\e}{\mathrm{e}}

\newcommand{\Gr}{\ensuremath{\mathrm{Gr}}}

\renewcommand{\roman}[1]{\textsc{\MakeLowercase{#1}}}

\newcommand*\chem[1]{\ensuremath{\mathrm{#1}}}
\newcommand{\pos}[1]{\ensuremath{\mathrm{#1}^+}}

\renewcommand{\neg}[1]{\ensuremath{\mathrm{#1}^-}}

\newcommand{\tctext}[1]{\begin{CJK}{UTF8}{bkai}#1\ignorespacesafterend\end{CJK}}